\documentclass[aps, amsmath,amssymb,prb,twocolumn,superscriptaddress]{revtex4-2}
\usepackage{graphicx}
\usepackage[colorlinks,bookmarks=false,citecolor=blue,linkcolor=red,urlcolor=blue]{hyperref}
\usepackage[utf8]{inputenc}
\usepackage{amsmath,bm}
\usepackage{bbold}
\usepackage{color,soul}
\usepackage{xstring}
\usepackage{wasysym}
\usepackage{xspace}
\usepackage{amssymb}
\usepackage{time}
\usepackage{bbold}
\usepackage{subfigure}
\usepackage{multirow}
\usepackage{xspace}
\usepackage{url}
\usepackage{amsmath}
\usepackage{booktabs}
\usepackage{lipsum}
\usepackage{cases}

\def\<{\langle}
\def\>{\rangle}

\usepackage{ulem}

\begin{document}

\title{Spectral $\mathcal{PT}$-symmetry breaking decoupled from thermodynamic criticality in a two-dimensional complex-conjugate Ising model}
\author{\firstname{Yihan} \surname{Wang}}
\email{These authors contributed equally to this work.}
\affiliation{\mbox{School of Physical Science and Technology, Beijing University of Posts and Telecommunications, Beijing 100876, China}}

\author{\firstname{Yuting} \surname{Wang}}
\email{These authors contributed equally to this work.}
\affiliation{\mbox{School of Physical Science and Technology, Beijing University of Posts and Telecommunications, Beijing 100876, China}}

\author{\firstname{Lihui } \surname{Sun}}
\affiliation{\mbox{School of Physical Science and Technology, Beijing University of Posts and Telecommunications, Beijing 100876, China}}

\author{\firstname{Qionglin} \surname{Dai}}
\affiliation{\mbox{School of Physical Science and Technology, Beijing University of Posts and Telecommunications, Beijing 100876, China}}
\affiliation{\mbox{Beijing Key Laboratory of Information Metamaterials, Beijing University of Posts and Telecommunications, Beijing 100876, China}}

\author{\firstname{Haihong} \surname{Li}}
\affiliation{\mbox{School of Physical Science and Technology, Beijing University of Posts and Telecommunications, Beijing 100876, China}}
\affiliation{\mbox{Beijing Key Laboratory of Information Metamaterials, Beijing University of Posts and Telecommunications, Beijing 100876, China}}

\author{\firstname{Lufeng} \surname{Zhang}}
\affiliation{\mbox{School of Physical Science and Technology, Beijing University of Posts and Telecommunications, Beijing 100876, China}}
\affiliation{\mbox{Beijing Key Laboratory of Information Metamaterials, Beijing University of Posts and Telecommunications, Beijing 100876, China}}

\author{\firstname{Yuhai} \surname{Liu}}
\email{yuhailiu@bupt.edu.cn}
\affiliation{\mbox{School of Physical Science and Technology, Beijing University of Posts and Telecommunications, Beijing 100876, China}}
\affiliation{\mbox{Beijing Key Laboratory of Information Metamaterials, Beijing University of Posts and Telecommunications, Beijing 100876, China}}
\date{\today}

\begin{abstract}
We study a two-dimensional anisotropic Ising model with complex-conjugate couplings. Analytic continuation of the exact anisotropic Ising critical condition yields a candidate thermodynamic critical line, which is independently supported by finite-size scaling consistent with the two-dimensional Ising universality class. Exact diagonalization of the non-Hermitian transfer matrix and finite-size extrapolation identify a distinct spectral boundary associated with $\mathcal{PT}$-symmetry breaking. For the parameter values examined, this boundary lies below the thermodynamic critical line, $K_{\rm PT}<K_c$, leaving an intermediate regime in which the dominant transfer-matrix spectrum is complex while the system remains thermodynamically disordered. The spatial correlations remain short-ranged at $K_{\rm PT}$, in contrast to the critical correlations at $K_c$, directly distinguishing the spectral transition from thermodynamic criticality. The complex-conjugate anisotropy modifies the finite-size free-energy amplitude, and accounting for the corresponding anisotropy factor recovers the Ising central charge $c=1/2$. These results demonstrate that spectral $\mathcal{PT}$-symmetry breaking can occur without thermodynamic criticality, establishing a decoupling between spectral and thermodynamic transitions in a two-dimensional non-Hermitian statistical system.
\end{abstract}
\maketitle
\section{Introduction}
\label{sec:intro}
Non-Hermiticity naturally arises in open quantum systems, where coupling to an environment can generate complex effective Hamiltonians and spectra~\cite{El-Ganainy2018}. Bender and Boettcher showed that a non-Hermitian Hamiltonian invariant under combined parity and time-reversal ($\mathcal{PT}$) symmetry can possess an entirely real spectrum in the $\mathcal{PT}$-unbroken regime~\cite{PhysRevLett.80.5243, Bender_2007}.
Experimental realizations of $\mathcal{PT}$-symmetric systems have been reported in photonics~\cite{Ruter2010}, acoustics~\cite{Shi2016}, ultracold atoms~\cite{Li2019}, and superconducting quantum processors~\cite{Dogra2021}. Of particular relevance here is $\mathcal{PT}$-symmetry breaking, a spectral transition in which real eigenvalues become complex-conjugate pairs.

Non-Hermitian concepts also extend naturally to classical statistical mechanics through the transfer-matrix formulation, which establishes a correspondence between classical statistical systems and quantum Hamiltonians. Hatano and Nelson demonstrated this connection through a path-integral formulation~\cite{PhysRevLett.77.570}. In the presence of an appropriate generalized $\mathcal{PT}$ symmetry, classical systems with complex Boltzmann weights can possess real partition functions despite having non-Hermitian transfer matrices~\cite{Meisinger2011, Meisinger2013}. Such systems can exhibit $\mathcal{PT}$-symmetry breaking, with complex-conjugate subleading transfer-matrix eigenvalues potentially giving rise to oscillatory spatial correlations~\cite{Meisinger2011, Meisinger2013}. Related non-Hermitian and complex-weight phenomena have also been investigated in two-dimensional classical spin models with imaginary magnetic fields or complex interactions~\cite{Zhang2020, Nishiyama2020, Krcmar2022}. A basic question, however, remains: does $\mathcal{PT}$-symmetry breaking of a classical transfer-matrix spectrum necessarily signal thermodynamic criticality, or can the two transitions occur at distinct points?

Here we address this question using an anisotropic square-lattice Ising model with complex-conjugate couplings, $K_x=K+i\phi$ and $K_y=K-i\phi$, which we call the complex-conjugate-coupled Ising (CCI) model. The construction is motivated in part by the generalized Baxter--Wu model with complex-conjugate couplings~\cite{blote2017,ffsy-k96l,wangyt-CBW-2026}, whose partition function is real despite its complex Boltzmann weights. Its transfer matrix, however, is Hermitian~\cite{wangyt-CBW-2026}, precluding the spectral $\mathcal{PT}$-symmetry breaking of interest here. Previous studies of the anisotropic two-dimensional Ising model at complex temperature have revealed complex spectra and spatially modulated correlations~\cite{PhysRevResearch.4.013018}. Unlike the complex-temperature setting considered there, the complex-conjugate couplings in the CCI model give rise to a non-Hermitian but $\mathcal{PT}$-symmetric setting, which can be used to examine the relation between spectral symmetry breaking and thermodynamic criticality.

The complex Boltzmann weights of the CCI model pose a sign problem. We investigate the model using complementary numerical approaches: reweighting Monte Carlo with configuration pairing~\cite{PhysRevB.41.9301,wangyt-CBW-2026} and Wang--Landau sampling~\cite{PhysRevLett.86.2050,PhysRevE.64.056101} to probe its thermodynamic behavior, together with exact diagonalization of the non-Hermitian transfer matrix to characterize its spectral properties and finite-size free-energy scaling. Analytic continuation of the exact anisotropic Ising critical condition~\cite{PhysRev.65.117} yields a candidate critical line that is supported by Monte Carlo finite-size scaling over the numerically accessible range $\phi\in(0,0.4]$. The critical scaling is consistent with the two-dimensional Ising universality class, while the finite-size free-energy scaling yields the Ising central charge $c=1/2$~\cite{Cardy1996}. In contrast, exact diagonalization reveals the emergence of complex dominant transfer-matrix eigenvalues. Finite-size extrapolation of the spectral crossings identifies a distinct boundary $K_{\rm PT}$, associated with $\mathcal{PT}$-symmetry breaking, that lies on the disordered side of the thermodynamic critical line $K_c$. The two boundaries enclose an intermediate regime in which the dominant transfer-matrix spectrum is complex while the system remains thermodynamically disordered. The contrasting spatial correlation behavior at $K_{\rm PT}$
and $K_c$ further supports the separation of the two transitions. Related separations between conventional criticality and spectral $\mathcal{PT}$ transitions have been reported in non-Hermitian quantum Ising chains~\cite{STARKOV2023169268,PhysRevB.110.014441}; here we establish this decoupling in a two-dimensional classical statistical model.

The remainder of this paper is organized as follows. Sec.~\ref{sec:model} introduces the CCI model and the numerical methods. Sec.~\ref{sec:Thermodynamic} investigates thermodynamic criticality and the associated universality class. Sec.~\ref{sec:PT} examines the spectral $\mathcal{PT}$-symmetry-breaking boundary and spatial spin correlations. Finally, Sec.~\ref{sec:conclusions} presents our conclusions and outlook.

\section{Model and methods}
\label{sec:model}
\begin{figure}[htbp]
\centering
\includegraphics[width=0.45\textwidth]{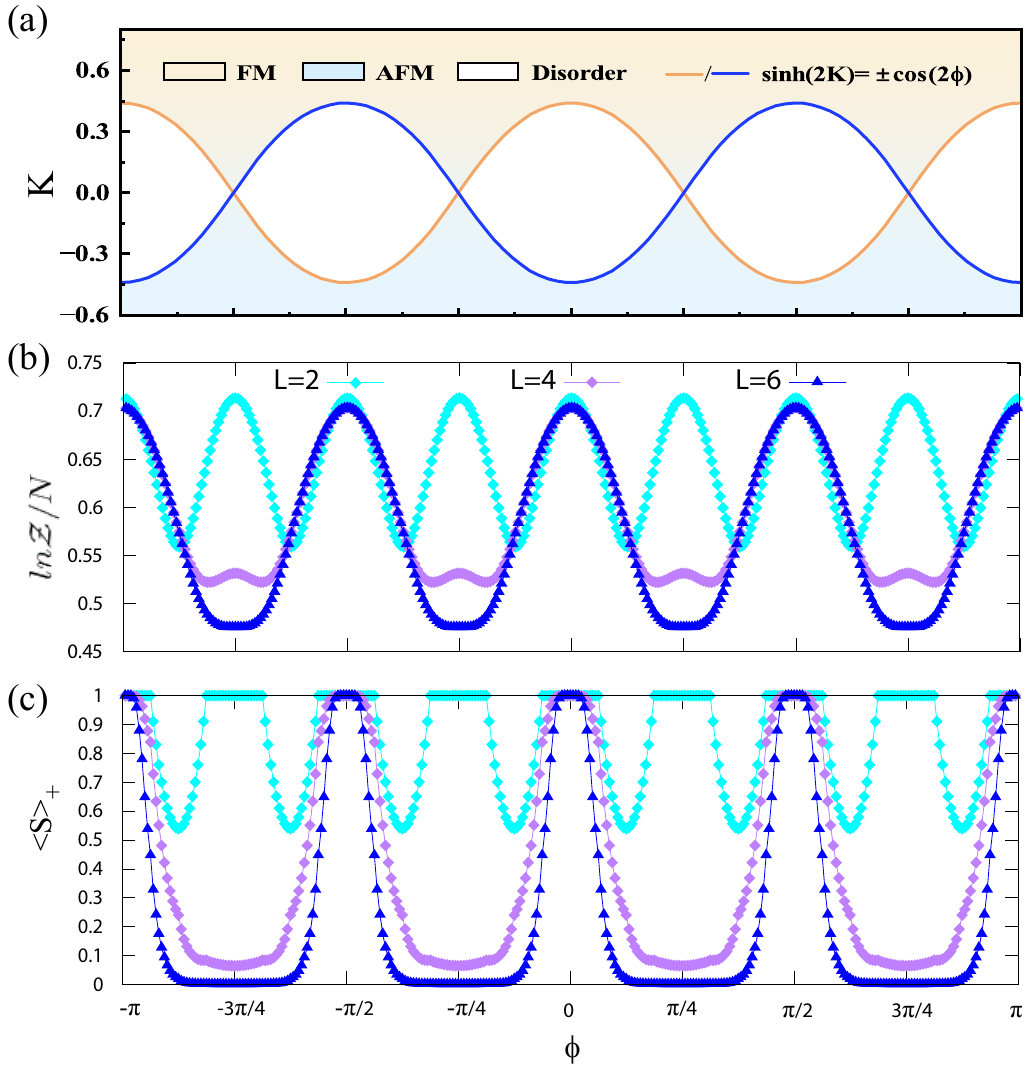}
\caption{\label{fig:phase_diagram}
Candidate phase structure, partition function, and average sign of the CCI model. (a) Analytically continued candidate phase boundaries from Eq.~\eqref{Eq:critical_line2}. The shaded regions indicate the ferromagnetic (FM), antiferromagnetic (AFM), and disordered phases.
(b) $(\ln\mathcal{Z})/N$ and (c) average sign $\langle S\rangle_+$ as functions of $\phi$ at $K=0.1$ for $L=2$, $4$, and $6$, obtained by exact enumeration. Both exhibit a period of $\pi/2$ in $\phi$.}
\end{figure}

We consider the anisotropic Ising model with the reduced Hamiltonian
\begin{equation}
\label{Ea:Hamiltonian1}
    \beta H= -K_x \sum_{\langle ij\rangle_x}\sigma_i\sigma_j - K_y \sum_{\langle mn \rangle_y}\sigma_m\sigma_n,
\end{equation}
where $\sigma_i=\pm1$, $\beta=1/k_BT$, and $K_x$, $K_y$ denote the reduced couplings along the horizontal and
vertical lattice directions, respectively. For real $K_x$ and $K_y$, the exact critical line of the anisotropic Ising model is given by Onsager's solution~\cite{PhysRev.65.117} as $\sinh(2K_x)\sinh(2K_y) = 1$, consistent with the Kramers-Wannier self-duality~\cite{PhysRev.60.252}.
For the CCI model, we set $K_x=K+i\phi$ and $K_y=K-i\phi$, where $K$ and $\phi$ are real. Substitution into the above condition gives
\begin{equation}
\label{Eq:critical_line2}
\sinh(2K) = \pm \cos(2\phi),
\end{equation}
which defines candidate boundaries separating the disordered phase from the ferromagnetic ($K>0$) and antiferromagnetic ($K<0$) ordered phases, as shown in Fig.~\ref{fig:phase_diagram}(a). In the following Monte Carlo investigation, we focus on the $K>0$ region and test whether the analytically continued line indeed corresponds to the genuine thermodynamic critical boundary.

\begin{figure}[htbp]
\centering
\includegraphics[width=0.45\textwidth]{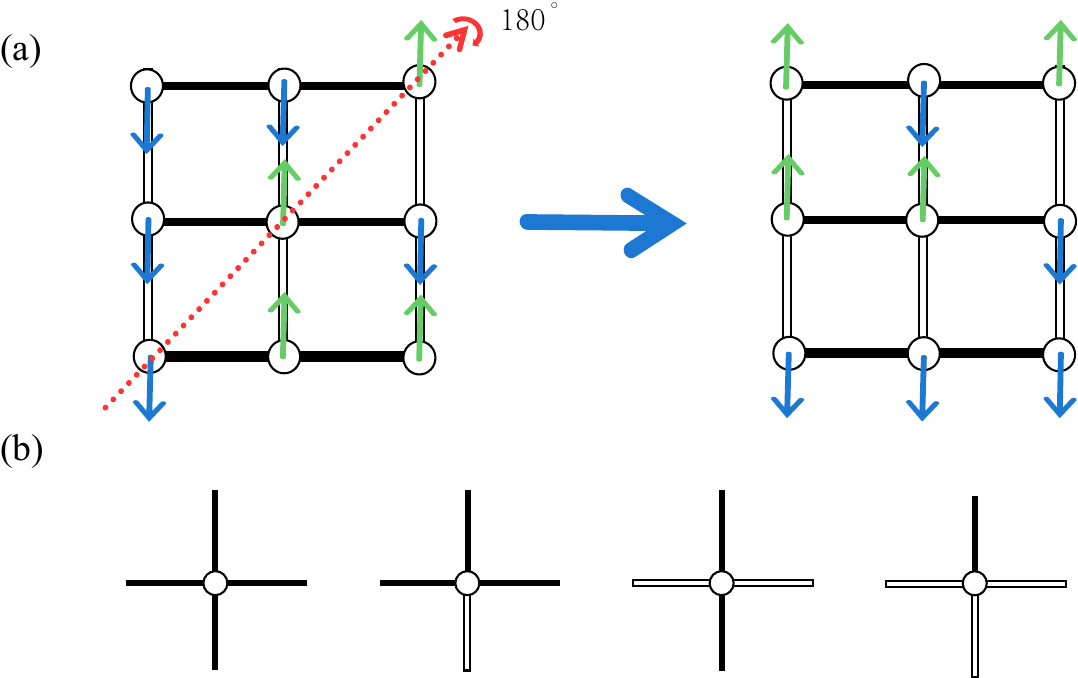}
\caption{\label{fig:lattice}
Configuration pairing and local bond structure of the CCI model. (a) A configuration $\Gamma$ and its counterpart $\Gamma'$ related by a $180^\circ$ rotation about a lattice-diagonal axis, which interchanges the horizontal and vertical bond sums. (b) Four types of local bond configurations around a central spin. Flipping the central spin changes $H_{i\phi}$ by $0$, $4$, $8$, or $0$ in magnitude, respectively. Solid and hollow bonds represent antiparallel and parallel spins, respectively.}
\end{figure}

The complex-conjugate couplings give rise to complex Boltzmann weights, which prevent the direct application of conventional importance-sampling Monte Carlo methods. To obtain a real-weight representation, we pair each configuration $\Gamma$ with its counterpart $\Gamma'$, obtained by a $180^\circ$ rotation about a lattice-diagonal axis, as illustrated in Fig.~\ref{fig:lattice}(a). The combined Boltzmann weight is then given by
\begin{equation}
\label{Eq:bundled_weight}
    W(\widetilde{\Gamma}) = 2e^{K H_K(\Gamma)}\cos\left[\phi H_{i\phi}(\Gamma)\right],
\end{equation}
where $H_K(\Gamma)$ and $H_{i\phi}(\Gamma)$ denote the sum and difference of the horizontal and vertical bond-spin products, respectively:
\begin{equation}
    H_K(\Gamma) \equiv -\frac{\partial(\beta H)}{\partial K}
= \sum_{\langle ij\rangle_x} \sigma_i\sigma_j + \sum_{\langle mn \rangle_y} \sigma_m\sigma_n,
\end{equation}
\begin{equation}
    H_{i\phi}(\Gamma) \equiv -\frac{\partial(\beta H)}{\partial(i\phi)}
= \sum_{\langle ij\rangle_x} \sigma_i\sigma_j - \sum_{\langle mn \rangle_y} \sigma_m\sigma_n.
\end{equation}
The configuration pairing eliminates the imaginary part of the Boltzmann weight, while the remaining cosine factor makes the sign problem explicit, since it can become negative.

The partition function and the positive-weight reference partition function are defined as
\begin{equation}
\label{abPf}
\mathcal{Z}=\frac{1}{2}\sum_{\Gamma}W(\widetilde{\Gamma}),
\qquad
\mathcal{Z}_{+}=\frac{1}{2}\sum_{\Gamma}|W(\widetilde{\Gamma})|.
\end{equation}
Here, the sums run over the original spin configurations,
and the factor $1/2$ compensates for double counting
in the configuration-pairing representation.
The average sign is then given by
\begin{equation}
\langle S\rangle_+
=\frac{\sum_{\Gamma}W(\widetilde{\Gamma})}
{\sum_{\Gamma}|W(\widetilde{\Gamma})|}
=\frac{\mathcal{Z}}{\mathcal{Z}_{+}},
\end{equation}
where $S=\mathrm{sgn}[W(\widetilde{\Gamma})]$ and
$\langle\cdots\rangle_+$ denotes an average over
the positive-weight reference ensemble.

The pairing construction also reveals a periodicity in the CCI model. As shown in Fig.~\ref{fig:lattice}(b), a single-spin flip changes $H_{i\phi}$ by an integer multiple of $4$. Since $H_{i\phi}=0$ for a fully aligned spin configuration, $H_{i\phi}$ is always a multiple of $4$. Consequently, the paired Boltzmann weights are periodic in $\phi$ with period $\pi/2$.

This periodicity is directly inherited by both the partition function [Fig.~\ref{fig:phase_diagram}(b)] and the average sign [ Fig.~\ref{fig:phase_diagram}(c)]. It is also consistent with the analytically continued critical condition in Eq.~\eqref{Eq:critical_line2}, whose two branches are interchanged under $\phi\rightarrow\phi+\pi/2$. Unlike in the generalized Baxter--Wu model, where similar periodicity generates additional self-dual lines~\cite{wangyt-CBW-2026}, no additional branches of the analytically continued critical condition arise in the CCI model.

To investigate the thermodynamic properties of the CCI model, we employ Monte Carlo simulations with sign reweighting~\cite{PhysRevB.41.9301,wangyt-CBW-2026}. Configurations are sampled according to the positive reference weight $|W(\widetilde{\Gamma})|$, and expectation values of observables $O$ are obtained from $\langle O\rangle
=\langle \widetilde O S\rangle_+/\langle S\rangle_+$, where $\widetilde O$ denotes the observable defined for a paired configuration. Our Monte Carlo simulations are performed in the $K>0$ region for $\phi\in(0,0.4]$. Details of the construction of paired observables and the Monte Carlo implementation are provided in Appendix~\ref{reweighting}.

As a complementary approach, we employ Wang--Landau sampling~\cite{PhysRevLett.86.2050,PhysRevE.64.056101}, which enables the detection of thermodynamic phase transitions without prior knowledge of an order parameter. By evaluating the second derivative of the free energy with respect to $K$, this method provides an independent thermodynamic check of the sign-reweighting results. The Wang--Landau calculations are performed for systems up to $L=18$, using overlapping sampling windows~\cite{PhysRevLett.110.210603}. Further numerical details are provided in Appendix~\ref{WL}.

The complex-conjugate couplings give rise to a non-Hermitian but $\mathcal{PT}$-symmetric model. This symmetry requires a spatial reflection $\mathcal{P}$ that interchanges the horizontal and vertical directions, combined with time reversal $\mathcal{T}$ (complex conjugation). Although this spatial reflection is not an exact symmetry of the finite-width strip geometry, the corresponding symmetry is expected to be restored in the thermodynamic limit. By exactly diagonalizing the transfer matrix for various finite widths, we determine finite-size spectral crossing points and extrapolate them to the thermodynamic limit $L_x \to \infty$ to estimate the spectral boundary associated with $\mathcal{PT}$-symmetry breaking. This allows us to examine the relation between the spectral transition and thermodynamic criticality. An approximate mapping based on the transfer-matrix formulation also connects the two-dimensional CCI model to a one-dimensional non-Hermitian quantum Ising chain. Details of the transfer-matrix formulation and the approximate quantum mapping are provided in Appendix~\ref{TMSP}.
\section{Thermodynamic criticality}
\label{sec:Thermodynamic}
Whether the candidate phase boundaries obtained by analytic continuation correspond to genuine thermodynamic transitions requires numerical verification. Although the CCI Hamiltonian preserves the global spin-flip $\mathbb{Z}_2$ symmetry, this symmetry alone does not uniquely determine the universality class, as illustrated by tricritical Ising behavior ~\cite{PhysRevB.33.7700,Mandal2016}. The persistence of two-dimensional Ising critical behavior for $\phi\neq0$ therefore needs to be established independently.

\begin{figure}
    \centering  \includegraphics[width=0.45\textwidth]{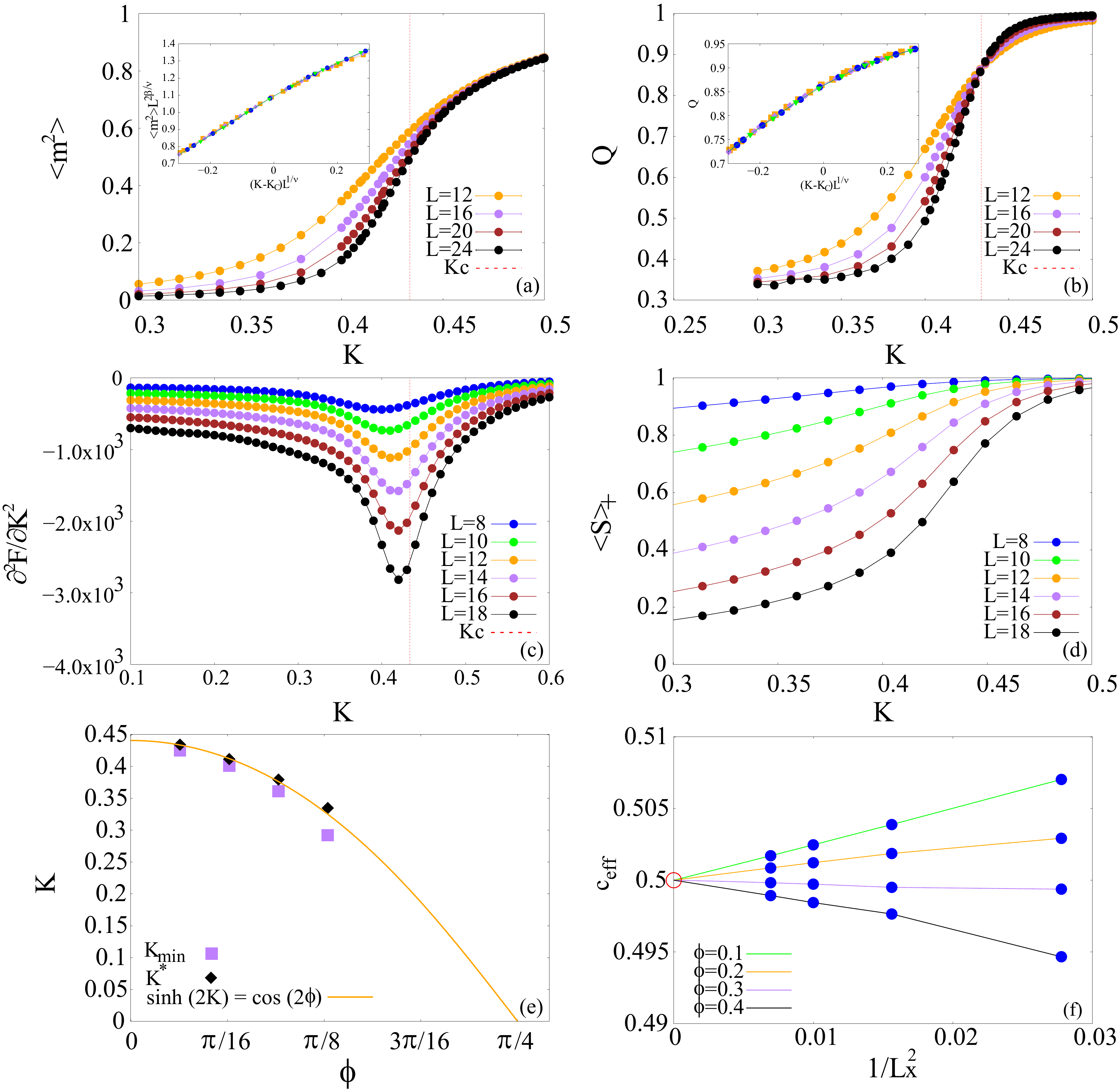}
    \caption{\label{z model-reweighting}
    Thermodynamic criticality and finite-size scaling of the CCI model. (a) Squared magnetization $\langle m^2 \rangle$ and (b) Binder ratio $Q$ versus $K$ for different system sizes at $\phi=0.1$. The red dashed lines indicate the analytically continued critical coupling $K_c\simeq0.4336$, and the insets show data collapses using the two-dimensional Ising critical exponents. (c) Second derivative of the free energy, $\partial^2F/\partial K^2$, obtained from Wang--Landau sampling, and (d) average sign $\langle S\rangle_+$ versus $K$ for different system sizes at $\phi=0.1$. (e) Transition estimates $K^*$ from Binder-ratio crossings for $(L_1,L_2)=(20,24)$ and $K_{\min}$ from the minimum of $\partial^2F/\partial K^2$ at $L=18$, compared with the analytically continued critical line. (f) Anisotropy-corrected effective central charge $c_{\rm eff}(L_x)$ versus $1/L_x^2$ for $\phi=0.1, 0.2, 0.3$ and $0.4$. Symbols denote transfer-matrix results, solid lines the exact free-fermion predictions, and the open circle the universal Ising central charge $c=1/2$.}
\end{figure}

We characterize the ferromagnetic transition using the squared magnetization
$m^2=\left(\frac{1}{N}\sum_{i=1}^{N}\sigma_{i}\right)^2$ and the Binder ratio $Q=\langle m^2\rangle^2/\langle m^4\rangle.$ This ratio is a dimensionless quantity that becomes scale invariant at criticality, allowing the critical point to be identified from finite-size crossings. Fig~\ref{z model-reweighting}(a) and \ref{z model-reweighting}(b) show $\langle m^2\rangle$ and $Q$, respectively, as functions of $K$ at $\phi=0.1$. As $L$ increases, the $\langle m^2\rangle$ rises more sharply near the candidate critical coupling, while the Binder-ratio curves cross near the same point. According to the finite-size scaling hypothesis~\cite{Luck1985}, the squared magnetization and Binder ratio satisfy
$\langle m^2\rangle=L^{-2\beta/\nu}\mathcal{M}\left[(K-K_c)L^{1/\nu}\right]$ and $Q=\mathcal{Q}\left[(K-K_c)L^{1/\nu}\right]$,
where $\beta$ and $\nu$ are the order-parameter and correlation-length critical exponents, respectively. Using the two-dimensional Ising values $\beta=1/8$ and $\nu=1$, together with the analytically continued coupling $K_c=\operatorname{arsinh}[\cos(0.2)]/2\simeq0.4336$, we obtain the data collapses shown in the insets of Figs.~\ref{z model-reweighting}(a) and \ref{z model-reweighting}(b). The agreement supports a thermodynamic transition at the predicted coupling with critical behavior consistent with the two-dimensional Ising universality class.

Although the order-parameter analysis identifies the thermodynamic transition, it cannot exclude additional phases or transitions. We therefore perform an independent Wang--Landau analysis of the free energy. The accessible energy-pair configurations and the estimated joint density of states are presented in Fig.~\ref{fig:accessible_pairs} of Appendix~\ref{WL}. As shown in Fig~\ref{z model-reweighting}(c), at $\phi=0.1$, the second derivative of the free energy exhibits a negative peak near the analytically continued phase transition line. The calculation of $\partial^2F/\partial K^2$ is detailed in Appendix ~\ref{WL}, with the explicit expression given in Eq.~\eqref{eq:F_second_derivative}. As the system size increases, the position of the minimum shifts toward the transition point $K_c\approx 0.4336$. Scanning the parameter range $K \in [0, 0.6]$, we detect no signatures of additional phases or thermodynamic phase transitions, consistent with the analytically continued phase diagram. However, the average sign decreases rapidly at smaller $K$, as shown in Fig.~\ref{z model-reweighting}(d), making reliable simulations at larger system sizes difficult and limiting further finite-size verification in this regime.

Fig~\ref{z model-reweighting}(e) compares two independent numerical estimates of the thermodynamic transition with the analytically continued critical line given by Eq.~\eqref{Eq:critical_line2} . Both estimates agree with the predicted critical line within numerical accuracy, supporting its identification as the thermodynamic phase boundary. The average sign becomes strongly suppressed in certain regions of the parameter space, limiting the range accessible to reliable numerical calculations using both reweighting Monte Carlo and Wang--Landau sampling. Its dependence on $\phi$ is shown in
Fig.~\ref{fig:phase_diagram}(c).

The finite-size free energy provides a further probe of the critical theory. For a two-dimensional critical system with periodic boundary conditions, conformal field theory relates its leading finite-size correction to the central charge ~\cite{PhysRevLett.56.742,PhysRevLett.56.746}. In the CCI model, the complex anisotropy must be accounted for when extracting the universal central charge.

Using the exact free-fermion representation~\cite{RevModPhys.36.856} of the anisotropic Ising transfer matrix and analytically continuing to the complex-conjugate critical couplings, we obtain the thermodynamic-limit real free-energy density
\begin{equation}
\label{eq:f_infty}
\begin{aligned}
f_{R,\infty}=-\frac{1}{2}\ln|2\sinh(2K_x)|-\frac{1}{4\pi}\operatorname{Re}\int_0^{2\pi}\gamma(q)\,dq,
\end{aligned}
\end{equation}
where $\gamma(q)=2\operatorname{arsinh}[\sinh(2K_y)\sin(q/2)]$. The corresponding anisotropy-corrected finite-size effective central charge is

\begin{equation}
\label{eq:ceff_exact}
\begin{aligned}
c_{\rm eff}^{\rm exact}(L_x)=\frac{6L_x^2}{\pi\cos^2(2\phi)}\Bigg[
&\frac{1}{2L_x}\sum_{r=0}^{L_x-1}\operatorname{Re}\gamma(q_r)\\
&-\frac{1}{4\pi}\operatorname{Re}\int_0^{2\pi}\gamma(q)\,dq\Bigg],
\end{aligned}
\end{equation}
where $q_r=(2r+1)\pi/L_x$. The derivation and analytic branches are detailed in Appendix~\ref{C-c}.

The asymptotic expansion of Eq.~\eqref{eq:ceff_exact} yields $c_{\rm eff}^{\rm exact}(L_x)=c+O(L_x^{-2})$, where $c$ is the universal central charge. The factor $\cos^2(2\phi)$ corrects the anisotropy-dependent finite-size Casimir amplitude, which would otherwise approach $c\cos^2(2\phi)$ under the conventional isotropic-cylinder normalization. To verify this prediction, we calculate the real free-energy density $f_R(L_x)=-L_x^{-1}\ln|\lambda_{max}(L_x)|$ from the largest-modulus transfer-matrix eigenvalue $\lambda_{max}(L_x)$. Combining this numerical result with the exact thermodynamic-limit free-energy density $f_{R,\infty}$ given by Eq.~\eqref{eq:f_infty}, we obtain the anisotropy-corrected effective central charge
\begin{equation}
\label{eq:ceff}
c_{\rm eff}(L_x)=\frac{6L_x^2}{\pi\cos^2(2\phi)} [f_{R,\infty}-f_R(L_x)]
\end{equation}

Fig.~\ref{z model-reweighting}(f) compares the transfer-matrix results (symbols) with the exact free-fermion predictions from Eq.~\eqref{eq:ceff_exact} (solid lines). For $\phi=0.1,0.2,0.3$, and $0.4$, the results approach the common limit $c=1/2$ as $L_x\rightarrow\infty$, confirming the two-dimensional Ising universality class.

Taken together, the reweighting and Wang--Landau results support the analytically continued thermodynamic critical
line within the numerically accessible regime. The critical scaling and anisotropy-corrected central charge $c=1/2$ establish the two-dimensional Ising universality of the thermodynamic transition. Having identified the thermodynamic critical line, we now examine whether spectral $\mathcal{PT}$-symmetry breaking occurs at the same boundary.

\section{Spectral $\mathcal{PT}$-symmetry breaking}
\label{sec:PT}
\begin{figure}
    \centering  \includegraphics[width=0.5\textwidth]{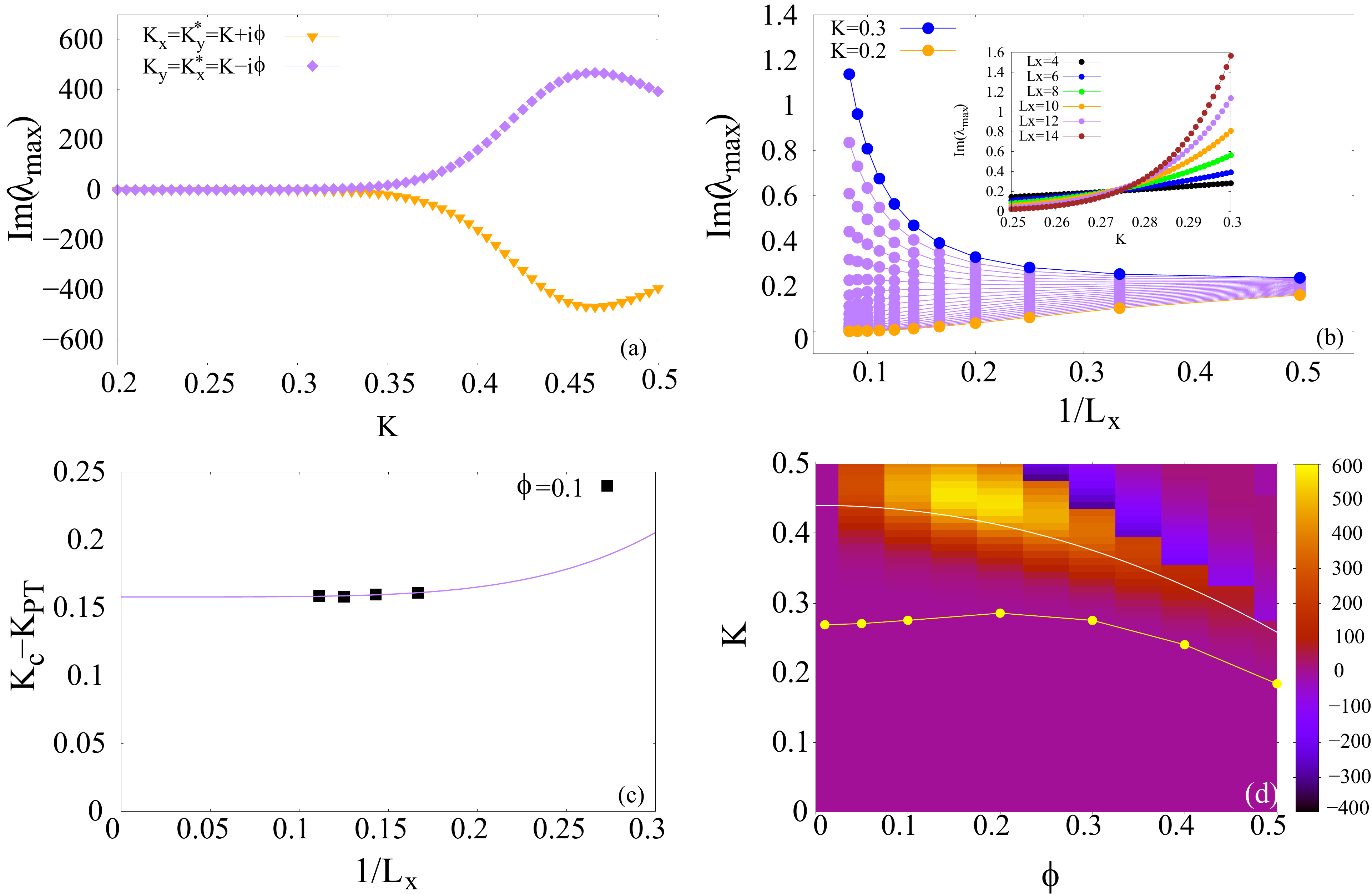}
    \caption{\label{fig:pt_boundary}Spectral transition and its separation from thermodynamic criticality. (a) Imaginary parts of the dominant transfer-matrix eigenvalues for the original and exchanged complex-conjugate couplings at $\phi=0.1$. The corresponding eigenvalues are complex conjugates. (b) Finite-size dependence of $\operatorname{Im}(\lambda_{\max})$ versus $1/L_x$ for different $K$ at $\phi=0.1$. Inset: Crossings of $\operatorname{Im}(\lambda_{\max})$ versus $K$ for different system sizes. (c) Finite-size extrapolation of $K_c-K_{\rm PT}$ at $\phi=0.1$, yielding $K_c-K_{\mathrm{PT}}\approx0.158$ in the thermodynamic limit. (d) $\operatorname{Im}(\lambda_{\max})$ in the $(\phi,K)$ plane for $L_x=12$. Yellow symbols and the white line denote $K_{\rm PT}$ and $K_c$, respectively.}
\end{figure}

We now examine the spectral properties of the non-Hermitian transfer matrix and their relation to thermodynamic criticality. The full eigenvalue spectrum for $L_x=4$, presented in Fig.~\ref{fig:S4}, illustrates the emergence of complex eigenvalues in different spectral branches. Here we focus on the dominant eigenvalue, which determines the thermodynamic free energy. As shown in Fig.~\ref{fig:pt_boundary}(a), exchanging the complex-conjugate couplings, $K_x\leftrightarrow K_y$, maps the dominant transfer-matrix eigenvalues to their complex conjugates, leaving the spectral boundary unchanged. This algebraic property, $T(K_y, K_x) = T^*(K_x, K_y)$, holds for any finite size. Although the finite strip geometry prevents a strict
finite-size $\mathcal{PT}$ symmetry, this complex-conjugation property is consistent with the expected restoration of $\mathcal{PT}$ symmetry in the thermodynamic limit.

To locate the spectral boundary, we examine the finite-size dependence of $\operatorname{Im}(\lambda_{\max})$. Fig.~\ref{fig:pt_boundary}(b) shows its evolution with $1/L_x$ for different values of $K$, revealing distinct and, in some cases, nonmonotonic finite-size behavior. The inset shows $\operatorname{Im}(\lambda_{\max})$ as a function of $K$ for different system sizes, with characteristic crossings near the spectral boundary. Using the crossing points obtained from system-size pairs $(L_x,L_x+5)$, we analyze the finite-size behavior of the
spectral boundary. As shown in Fig.~\ref{fig:pt_boundary}(c), a power-law fit of the form $K_{\mathrm{PT}}(L_x)=K_{\mathrm{PT}}(\infty)+aL_x^{-p}$, using $L_x=6$--$9$ at $\phi=0.1$, yields $K_c-K_{\mathrm{PT}}\approx0.158$ and an effective exponent $p\approx4.63$. However, the crossing points exhibit nonmonotonic finite-size behavior, and the fitted parameters remain sensitive to the choice of system sizes. Therefore, this extrapolation should be regarded as a tentative estimate. A more precise determination of the spectral boundary in the thermodynamic limit requires access larger system sizes. Nevertheless, the numerical results provide evidence for a distinct spectral boundary associated with PT-symmetry breaking.

The corresponding spectral phase diagram is shown in Fig.~\ref{fig:pt_boundary}(d), where $\operatorname{Im}(\lambda_{\max})$ is plotted in the $(\phi,K)$ plane for $L_x=12$. For $\phi \in [0.01, 0.1]$, $K_{\rm PT}$ is estimated by extrapolation to the thermodynamic limit, whereas for $\phi \in [0.2, 0.5]$, it is estimated from crossings between $L_x=12$ and $L_x=14$. The complex, nonmonotonic finite-size behavior prevents reliable thermodynamic-limit extrapolation for these larger values of $\phi$. Nevertheless, the numerical estimates consistently place $K_{\rm PT}$ below the thermodynamic critical line $K_c$, indicating an intermediate regime with complex dominant transfer-matrix eigenvalues while the system remains thermodynamically disordered.

To further characterize the intermediate regime, we investigate the spatial spin-correlation function. Previous studies of $\mathcal{PT}$-symmetric statistical models have shown that complex transfer-matrix eigenvalues can give rise to oscillatory spatial correlations~\cite{Meisinger2011,Meisinger2013}. More generally, spatial correlations are not determined by the dominant transfer-matrix eigenvalue alone, but depend on the ratios of relevant eigenvalues to the dominant one, together with the corresponding matrix elements of the spin operator between the left and right eigenstates~\cite{Baxter1982}.

In the present model, spin correlations along the horizontal direction are generally complex, with both real and imaginary components. Nevertheless, as shown in Fig.~\ref{fig:S1} of Appendix~\ref{reweighting}, they exhibit no pronounced regular spatial oscillations, despite the presence of complex transfer-matrix eigenvalues. Instead, the correlations show predominantly
decaying behavior.

We therefore focus on the diagonal spin-correlation function along the pairing axis $i=j$, defined by

\begin{equation}
\langle G_d(r)\rangle=\frac{1}{L}\sum_j
\left\langle
\sigma_{j,j}\sigma_{j+r,j+r}
\right\rangle .
\label{eq:diagonal_corr}
\end{equation}
Here, $r$ denotes the separation in lattice steps along the diagonal axis. Unlike the horizontal and vertical correlations, this average is restricted to the pairing axis rather than all lattice translations.

Under the configuration-pairing transformation $\Gamma\rightarrow\Gamma'$, which exchanges the two lattice coordinates, the configuration-level diagonal correlation observable is invariant, $G_d(r,\Gamma')=G_d(r,\Gamma)$. Together with the complex-conjugate pairing of the corresponding Boltzmann weights, this invariance ensures that the expectation value $\langle G_d(r)\rangle$ remains real, even when the transfer-matrix spectrum contains complex eigenvalues. The detailed derivation and the corresponding reweighting procedure are given in Appendix~\ref{reweighting}.

\begin{figure}
\centering
\includegraphics[width=0.5\textwidth]{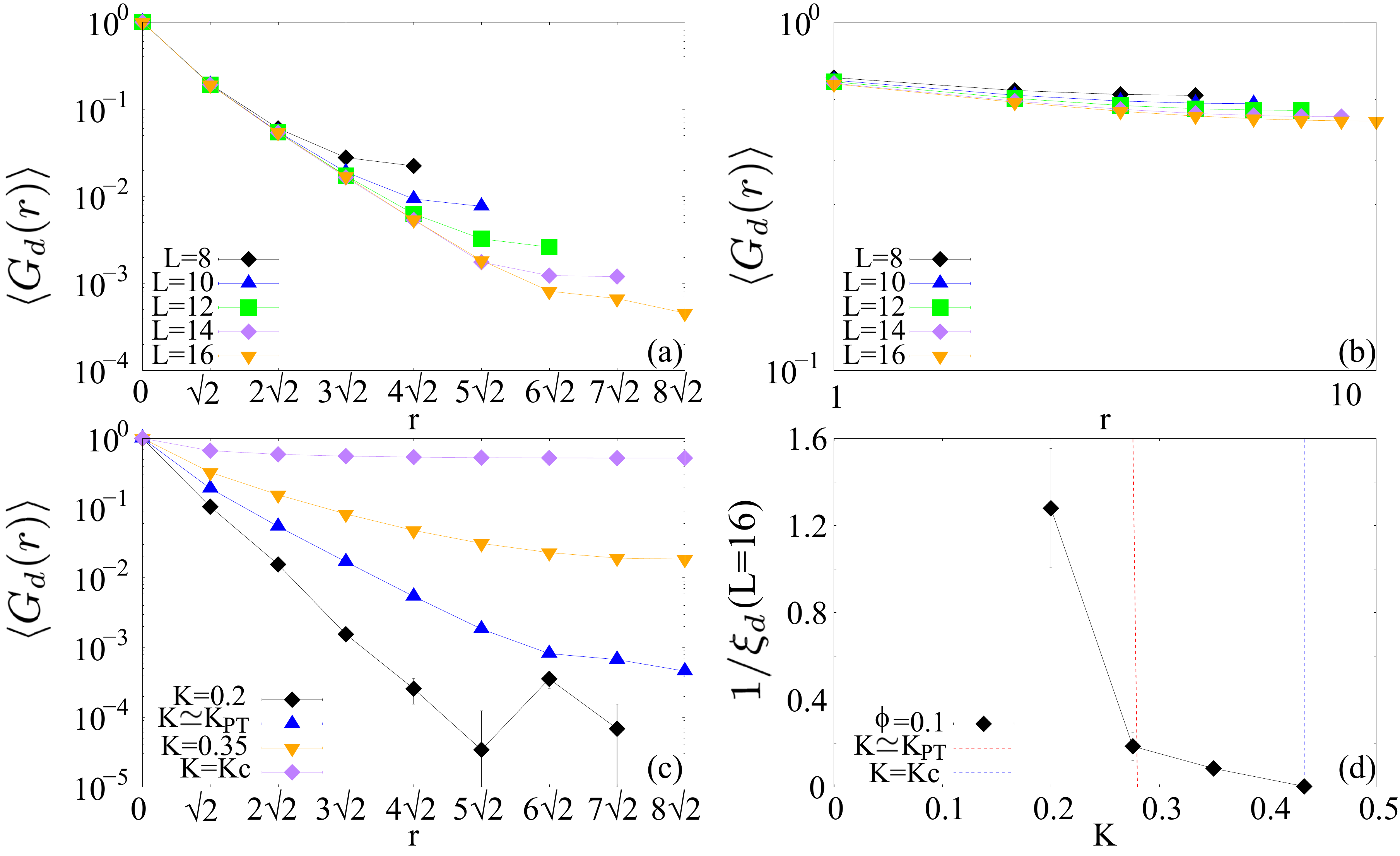}
\caption{\label{fig:correlation}Diagonal spin correlations across the spectral and thermodynamic boundaries at $\phi=0.1$. (a) Diagonal spin-correlation function $\langle G_d(r)\rangle$ at $K_{\rm PT}$ for different system sizes,
shown on a semilogarithmic scale to illustrate the approximately exponential decay. (b) Corresponding diagonal spin-correlation function at $K_c$, shown on a log-log scale to illustrate behavior consistent with power-law decay.
(c) Diagonal spin-correlation function $\langle G_d(r)\rangle$ versus distance $r$ for different values of $K$ at $L=16$.
(d) Effective inverse correlation length $1/\xi_d$ as a function of $K$, extracted from exponential fits to the finite-size data in (c). The red and blue dashed lines indicate $K_{\rm PT}$ and $K_c$, respectively.}
\end{figure}

Figs.~\ref{fig:correlation}(a) and (b) show the diagonal spin-correlation function $\langle G_d(r)\rangle$ for different system sizes at $K_{\mathrm{PT}}$ and $K_c$, respectively. At $K_{\mathrm{PT}}$, the approximately linear behavior in the semilogarithmic plot in Fig.~\ref{fig:correlation}(a) is consistent with exponential decay, indicating short-range correlations. In contrast, at $K_c$, the much slower decay in the log-log plot in
Fig.~\ref{fig:correlation}(b) is consistent with power-law correlations. These contrasting behaviors indicate that the spectral transition occurs in a short-range-correlated regime, distinct from thermodynamic criticality.

Fig.~\ref{fig:correlation}(c) further shows the evolution of $\langle G_d(r)\rangle$ with $K$ at $L=16$ and $\phi=0.1$. As $K$ increases from $K_{\mathrm{PT}}$ to $K_c$, the spatial decay becomes progressively slower, reflecting the enhancement of correlations upon approaching the thermodynamic critical point. To quantify the spatial decay, we extract the effective inverse correlation length $1/\xi_d$ by fitting the correlation function to the exponential form $\langle G_d(r)\rangle\sim A_d e^{-r/\xi_d}$ using the finite-size data in Fig.~\ref{fig:correlation}(c). As shown in Fig.~\ref{fig:correlation}(d), $1/\xi_d$ remains appreciably nonzero at $K_{\mathrm{PT}}$ and decreases toward the thermodynamic critical point $K_c$. Although this quantity represents a finite-size estimate rather than the correlation length in the strict thermodynamic limit, its nonzero value at $K_{\mathrm{PT}}$ is consistent with short-range correlations at the spectral transition.

Furthermore, the results presented in Appendix~\ref{reweighting} show that spin correl  ations along the horizontal direction also exhibit short-range behavior at $K_{\mathrm{PT}}$, despite their generally complex-valued nature. This observation suggests that the short-range correlation
behavior at the spectral boundary is not specific to the diagonal direction.

The combination of the finite separation between $K_{\rm PT}$ and \(K_c\) and the persistence of short-range correlations at $K_{\rm PT}$ establishes a decoupling between spectral $\mathcal{PT}$-symmetry breaking and thermodynamic criticality in the two-dimensional CCI model.

\section{Conclusions and Outlook}
\label{sec:conclusions}
In summary, we have investigated the thermodynamic and spectral properties of a two-dimensional Ising model with complex-conjugate couplings. The thermodynamic critical boundary is consistent with the analytic continuation of Onsager's exact solution, while finite-size scaling and the anisotropy-corrected central charge support the two-dimensional Ising universality class. Through finite-size analysis of the non-Hermitian transfer-matrix spectrum, we identify a distinct spectral boundary associated with $\mathcal{PT}$-symmetry breaking at $K_{\rm PT}<K_c$. Spatial correlations remain short ranged at $K_{\rm PT}$, in contrast to the critical power-law correlations at $K_c$. An approximate mapping to a one-dimensional non-Hermitian quantum Ising chain, derived in Appendix~\ref{TMSP}, further connects the classical statistical model to non-Hermitian quantum spectral physics. Collectively, these results establish a decoupling between spectral $\mathcal{PT}$-symmetry breaking and thermodynamic criticality, revealing an intermediate regime with complex dominant transfer-matrix eigenvalues
while the system remains thermodynamically disordered.

While the finite strip geometry prevents strict finite-size $\mathcal{PT}$ symmetry, this symmetry is expected to be
restored in the thermodynamic limit. Future studies could explore whether a generalized $\mathcal{PT}$ operator can be constructed that ensures exact symmetry at all finite sizes, providing a more rigorous framework for investigating the interplay between
spectral symmetry breaking and thermodynamic criticality. Our results provide a basis for further investigations of spectral transitions in non-Hermitian statistical systems.

\begin{acknowledgments}
We would like to thank Wenan Guo for the valuable discussions. Y. Liu acknowledges support from the National Natural Science Foundation of China under Grant Nos. 12305039 and 12574251, the Fundamental Research Funds for the Central Universities from the Beijing University of Posts and Telecommunications under Grant Nos. 2025JCTP08 and 2025AI4S01, as well as the Open Fund of the Key Laboratory of Multiscale Spin Physics (Ministry of Education), Beijing Normal University, under Grant No. SPIN2025K03.
\end{acknowledgments}
\bibliography{reference}

\appendix
\twocolumngrid

\section{Configuration Pairing, Monte Carlo Sampling, and Spatial Correlations}
\label{reweighting}
As described in the main text and illustrated in Fig.~\ref{fig:lattice}(a), each configuration
$\Gamma$ is paired with its counterpart $\Gamma'$ under a $180^\circ$ rotation about a lattice-diagonal axis.
We choose the diagonal $i=j$ as the pairing axis, so that $\sigma_{i,j}(\Gamma')=\sigma_{j,i}(\Gamma)$.
This transformation leaves $H_K$ unchanged and reverses the sign of $H_{i\phi}$, with
$H_K(\Gamma')=H_K(\Gamma)$ and
$H_{i\phi}(\Gamma')=-H_{i\phi}(\Gamma)$. The original Boltzmann weights are therefore
\begin{equation}
\begin{aligned}
w(\Gamma)&=e^{K H_K(\Gamma)}e^{i\phi H_{i\phi}(\Gamma)},\\
w(\Gamma')&=e^{K H_K(\Gamma)}e^{-i\phi H_{i\phi}(\Gamma)}.
\end{aligned}
\end{equation}

For a general observable $O$, the paired observable corresponding to the paired configuration $\widetilde{\Gamma}$  is defined as
\begin{equation}
\label{eq:paired_observable}
\begin{aligned}
\widetilde O(\widetilde{\Gamma})
&=\frac{O(\Gamma)w(\Gamma)
+O(\Gamma')w(\Gamma')}
{w(\Gamma)+w(\Gamma')}\\
&=\frac{O(\Gamma)+O(\Gamma')}{2}+i\frac{O(\Gamma)-O(\Gamma')}{2}
\tan[\phi H_{i\phi}(\Gamma)].
\end{aligned}
\end{equation}
For observables invariant under configuration pairing, i.e., $O(\Gamma')=O(\Gamma)$, this reduces to
$\widetilde O(\widetilde{\Gamma})=O(\Gamma)$. This is the case for the magnetization moments $m^2$ and $m^4$ used to construct the Binder ratio $Q$ in the main text.

The Monte Carlo simulations are performed on $L\times L$ square lattices with periodic boundary conditions in both directions. The positive-weight reference ensemble, defined by the reference partition function $\mathcal{Z}_{+}$ in Eq.~\eqref{abPf}, is simulated using single-spin Metropolis updates~\cite{Metropolis1953}, with the paired weight $W(\widetilde{\Gamma})$ given by Eq.~\eqref{Eq:bundled_weight}. At each Monte Carlo step, a randomly selected spin is flipped with probability
\begin{equation}
P_{\mathrm{acc}}(\Gamma\to\Gamma_{\mathrm{new}})
=\min\left\{1,\frac{|W(\widetilde{\Gamma}_{\mathrm{new}})|}
{|W(\widetilde{\Gamma})|}
\right\},
\end{equation}
where $\Gamma_{\mathrm{new}}$ denotes the configuration obtained by flipping the selected spin.

\begin{figure}
    \centering
    \includegraphics[width=0.5\textwidth]{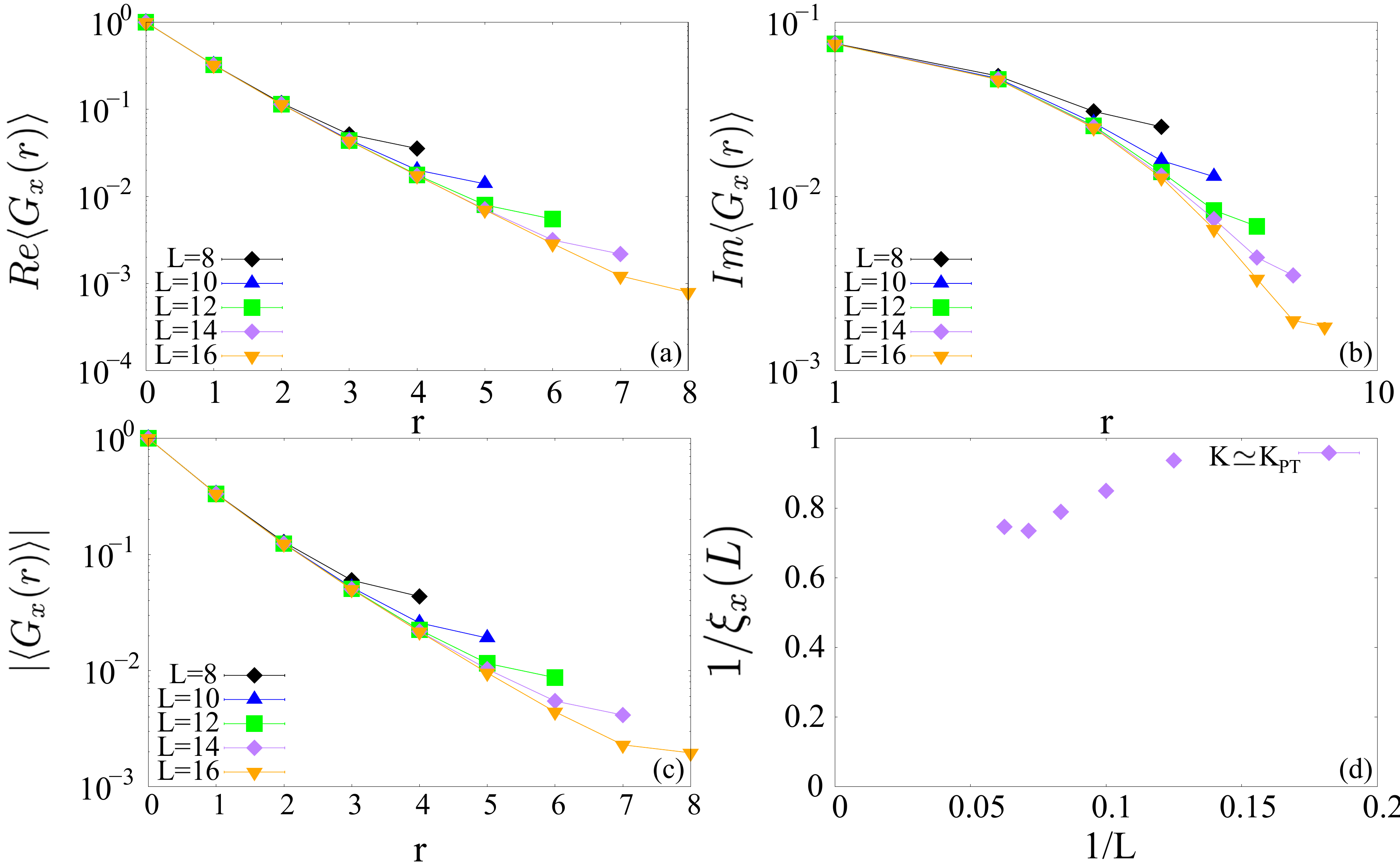}
    \caption{\label{fig:S1}{Spatial correlations along the horizontal direction
at $\phi=0.1$ and $K=K_{\rm PT}$. (a) Real and (b) imaginary parts of the reweighted horizontal spin-correlation function versus distance $r$ for different system sizes $L$. (c) Magnitude $|\langle G_x(r)\rangle|$ obtained from the real and imaginary parts in (a) and (b). (d) Inverse correlation length $1/\xi_x(L)$ extracted from the correlation magnitude as a function of $1/L$.}}
\end{figure}

For spatial correlations along the horizontal and vertical directions, we average over all lattice translations and define
\begin{equation}
\begin{aligned}
G_x(r,\Gamma)
&=\frac{1}{L^2}\sum_{i,j}
\sigma_{i,j}(\Gamma)
\sigma_{i+r,j}(\Gamma),\\
G_y(r,\Gamma)
&=\frac{1}{L^2}\sum_{i,j}
\sigma_{i,j}(\Gamma)
\sigma_{i,j+r}(\Gamma),
\end{aligned}
\end{equation}
Since the pairing transformation interchanges the lattice
indices $i$ and $j$, we have $G_x(r,\Gamma')=G_y(r,\Gamma)$.
The paired horizontal correlation function can therefore
be constructed using only the configuration $\Gamma$,

\begin{equation}
\label{eq:paired_correlation_x}
\begin{aligned}
\widetilde G_x(r,\widetilde{\Gamma})
&=\frac{G_x(r,\Gamma)+G_y(r,\Gamma)}{2}\\
&\quad+i\frac{G_x(r,\Gamma)-G_y(r,\Gamma)}{2}
\tan[\phi H_{i\phi}(\Gamma)].
\end{aligned}
\end{equation}
Its expectation value is evaluated using the sign-reweighting formula given in the main text,
$\langle G_x(r)\rangle=\langle \widetilde G_x(r)S\rangle_+/{\langle S\rangle_+}.$
The resulting correlation function is generally complex. Its magnitude is obtained by taking
the modulus after the reweighted average,
\begin{equation}
|\langle G_x(r)\rangle|=\sqrt{
[\operatorname{Re}\langle G_x(r)\rangle]^2
+[\operatorname{Im}\langle G_x(r)\rangle]^2
}.
\end{equation}
The horizontal spin correlations obtained from the above reweighting procedure are shown in Fig.~\ref{fig:S1}. At $\phi=0.1$ and $K=K_{\rm PT}$, Figs.~\ref{fig:S1}(a) and \ref{fig:S1}(b) display the real and imaginary parts of $\langle G_x(r)\rangle$ for different system sizes. The corresponding magnitudes, obtained by combining these two components, are shown in Fig.~\ref{fig:S1}(c). The correlation magnitude exhibits approximately exponential decay, from which we extract the finite-size correlation length $\xi_x(L)$.
The inverse correlation length $1/\xi_x(L)$, plotted against $1/L$ in Fig.~\ref{fig:S1}(d), decreases slowly with increasing system size and shows no indication of approaching zero over the accessible range of system sizes. These results provide complementary support for the short-range correlation behavior at the spectral boundary discussed in the main text.

To obtain a correlation observable invariant under the configuration-pairing transformation, we also measure spin correlations along the pairing axis $i=j$, as
defined in Eq.~\eqref{eq:diagonal_corr}. Since every site on this axis remains fixed under the pairing transformation, we have
$G_d(r,\Gamma')=G_d(r,\Gamma)$. Its expectation value is therefore given by $\langle G_d(r)\rangle =\langle G_d(r,\Gamma)S\rangle_+/\langle S\rangle_+$,
where $S$ is the reweighting factor defined above. Since $G_d(r,\Gamma)$ is real and invariant under configuration pairing, its expectation value is real, provided that the reweighted ensemble is well defined. Nevertheless, its numerical evaluation remains subject to sign cancellations in the reference ensemble. This invariance is not generally guaranteed for
diagonal correlations measured away from the pairing axis.

\section{Wang–Landau sampling}
\label{WL}
We use Wang--Landau (WL) sampling~\cite{PhysRevLett.86.2050,PhysRevE.64.056101} to estimate the joint density of states $g(H_K,H_{i\phi})$, which counts the original spin configurations with specified values of $H_K$ and $H_{i\phi}$. In terms of this density of states, the partition function can be reconstructed as
\begin{equation}
\label{ZPWL}
\mathcal{Z}(K,\phi)=
\sum_{(H_K,H_{i\phi})}
g(H_K,H_{i\phi})e^{KH_K}
\cos(\phi H_{i\phi}).
\end{equation}
This representation allows thermodynamic quantities to be reconstructed over different values of $K$ and $\phi$, providing an independent check of the thermodynamic transition identified in the main text.

\begin{figure}
    \centering
    \includegraphics[width=0.5\textwidth]{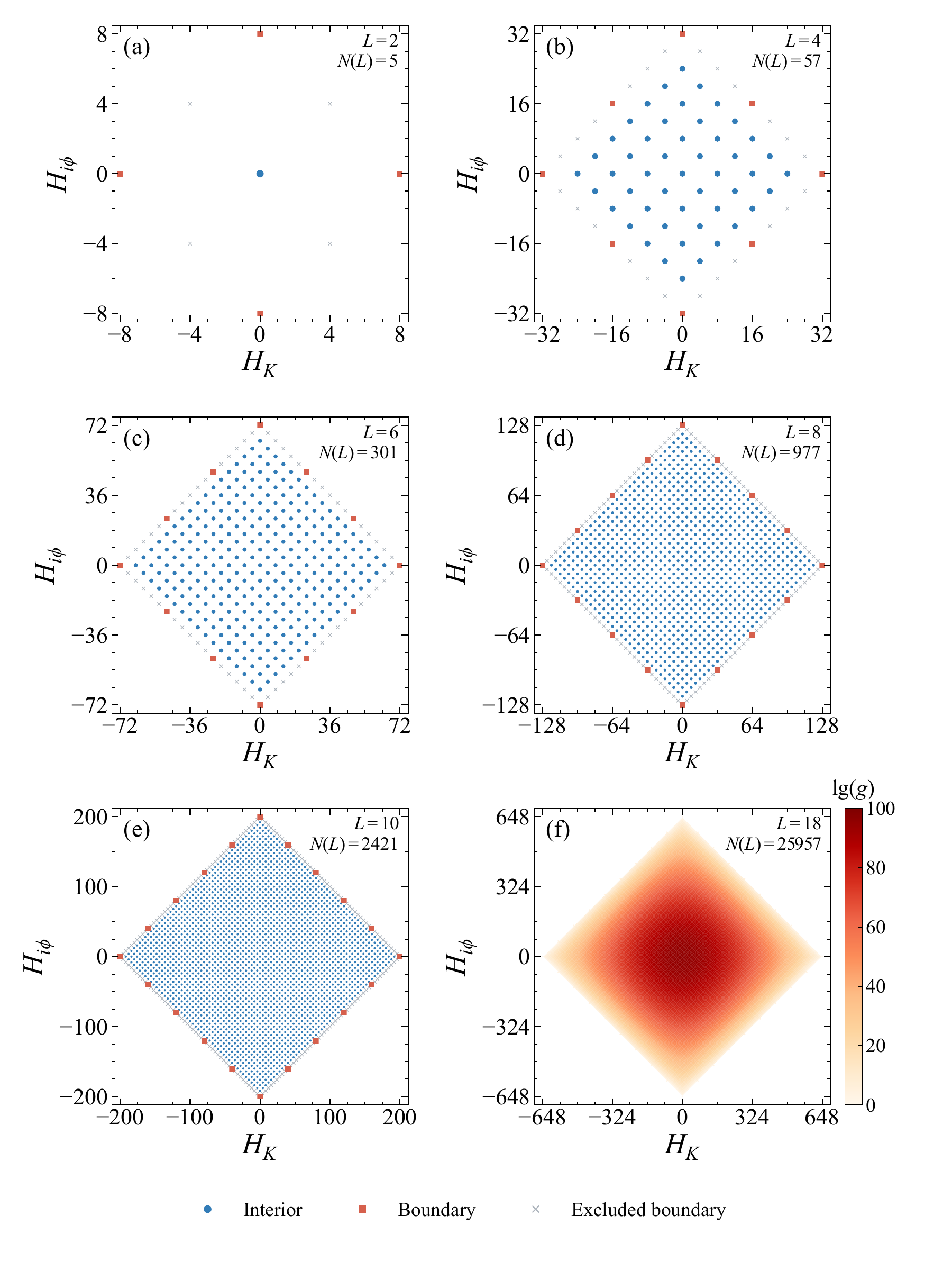}
    \caption{\label{fig:accessible_pairs}Accessible $(H_K,H_{i\phi})$ combinations and joint density of states of the CCI model. (a--e) Accessible combinations for $L=2$, $4$, $6$, $8$, and $10$, respectively. Blue circles and red squares denote accessible interior and boundary points, while gray crosses mark inaccessible boundary points. $N(L)$ gives the total number of accessible combinations. (f) Logarithm of the estimated joint density of states, $\lg g(H_K,H_{i\phi})$, for $L=18$. The color scale indicates the logarithmic density of states.}
\end{figure}

For an $L\times L$ square lattice with even $L$ and periodic boundary conditions in both directions, the accessible $(H_K,H_{i\phi})$ combinations form a diamond-shaped pattern, as shown in ~Figs.~\ref{fig:accessible_pairs}(a)--\ref{fig:accessible_pairs}(e). The numbers of accessible interior points in successive columns of increasing $H_K$ are $1,2,\ldots,L^2/2-1,\ldots,2,1$, giving $(L^2/2-1)^2$ interior points. On each boundary, all neighboring spins along one lattice direction are either parallel or antiparallel. Periodicity requires spin reversals along the other direction to occur in pairs, yielding $L/2+1$ accessible points on each edge. Counting each corner only once gives $2L$ distinct boundary points. The total number of accessible combinations is therefore
\begin{equation}
\label{eq:accessible_pair_count}
\begin{aligned}
N(L)=\left(\frac{L^2}{2}-1\right)^2+2L=\frac{L^4}{4}-L^2+2L+1.
\end{aligned}
\end{equation}

For system sizes extending beyond those illustrated in Figs.~\ref{fig:accessible_pairs}(a)--\ref{fig:accessible_pairs}(e), the formula gives $N(L)=5065, 9437, 16161, 25957$ for $L=12, 14, 16, 18$, respectively, consistent with the numbers of accessible $(H_K,H_{i\phi})$
combinations observed in our numerical sampling.
The rapid growth of $N(L)$ makes sampling the joint density of states increasingly demanding. We use overlapping windows following Refs.~\cite{PhysRevLett.110.210603,Vogel2014}. We assign each accessible $(H_K,H_{i\phi})$ combination an index $j$ and divide the indexed list into overlapping windows, within which WL sampling is performed separately. At each step, a randomly selected spin is flipped with probability
\begin{equation}
P_{\mathrm{acc}}(\Gamma\to\Gamma_{\mathrm{new}}) = \min\left\{ 1, \frac{ g[H_K(\Gamma),H_{i\phi}(\Gamma)] }{ g[H_K(\Gamma_{\mathrm{new}}), H_{i\phi}(\Gamma_{\mathrm{new}})] } \right\},
\end{equation}
where $g$ denotes the current estimate of the joint density of states within the window. A spin flip is rejected if the resulting energy pair lies outside the window. After each attempted flip, $\ln g$ is increased by $\ln f$ at the energy pair where the system resides, and the corresponding histogram count is incremented by one. The modification factor is initially set to $f=e$. The histogram is considered flat when every energy pair in the window has been visited at least $99\%$ of the mean number of visits. The modification factor is then reduced according to $f\to\sqrt{f}$, and the histogram is reset to zero for the next sampling iteration. This procedure is repeated until $\ln f\leq\ln f_{\min}$, with the prescribed threshold $\ln f_{\min}$ chosen between $10^{-8}$ and $10^{-6}$.

The density-of-states estimates obtained in different windows have independent normalization constants and must be aligned before being joined. Vogel et al. select joining points by matching the derivatives of $\ln g$ with respect to energy in the overlap region~\cite{Vogel2014}. We adapt this matching idea to the indexed list of energy pairs using the magnitudes of neighboring differences in $\ln g$. Writing the logarithmic density of states in window $a$ as $\ell_a(j)=\ln g_a(j)$, we calculate
\begin{equation}
d_a(j) = \left|\ell_a(j+1)-\ell_a(j)\right|.
\end{equation}
For adjacent windows $a$ and $a+1$, the joining point $j^*$ is chosen to minimize $ \left|d_a(j)-d_{a+1}(j)\right| $ where both $j$ and $j+1$ lie within their overlap region.
The logarithmic density of states in the latter window is then shifted by a constant,
\begin{equation}
\ell_{a+1}(j)\longrightarrow\ell_{a+1}(j)+C,
\end{equation}
where $C$ is chosen so that the shifted estimate agrees with the previously joined result at $j^*$. We retain the existing result up to and including $j^*$ and use the shifted estimate from window $a+1$ beyond this point. Repeating this procedure produces a relative density of states over the entire indexed list.

Our WL calculations reach system sizes up to $L=18$. Fig~\ref{fig:accessible_pairs}(f) displays the logarithm of the estimated joint density of states, $\lg g(H_K,H_{i\phi})$, for $L=18$. The density of states is concentrated in the central region of the $(H_K,H_{i\phi})$ plane and decreases toward the boundaries of the accessible energy space, reflecting the decreasing number of microscopic configurations associated with extreme energy values.Using Eq.~\eqref{ZPWL}, we calculate the second derivative of the dimensionless free energy $F=-\ln\mathcal{Z}$ with respect to $K$ at fixed $\phi$,
\begin{equation}
\label{eq:F_second_derivative}
\frac{\partial^2F}{\partial K^2} = -\frac{1}{\mathcal{Z}}
\frac{\partial^2\mathcal{Z}}{\partial K^2} + \frac{1}{\mathcal{Z}^2} \left( \frac{\partial\mathcal{Z}}{\partial K}
\right)^2,
\end{equation}
with
\begin{equation}
\begin{aligned}
\frac{\partial^n\mathcal{Z}}{\partial K^n} &= \sum_{(H_K,H_{i\phi})} g(H_K,H_{i\phi})H_K^n e^{K H_K}\cos(\phi H_{i\phi}), \\
&\hspace{4cm} n=1,2.
\end{aligned}
\end{equation}
The position of the minimum of the free-energy curvature provides a finite-size estimate of the transition point, which can be compared with the Binder-ratio crossings in the main text.

Although WL sampling estimates a nonnegative joint density of states, reconstructing the partition function involves
the sign-changing cosine factor in Eq.~\eqref{ZPWL}. As the average sign approaches zero, increasingly severe cancellations between contributions from different energy pairs amplify errors in the estimated density of states and limit the range of reliable numerical calculations. Thus, sampling with nonnegative weights does not by itself eliminate the exponential computational barrier
associated with the sign problem~\cite{wangyt-CBW-2026}.

\section{Transfer-matrix formulation and quantum mapping}
\label{TMSP}
\subsection{Transfer-matrix construction and spectral properties}
The partition function of the CCI  model on an $L_x\times L_y$ square lattice  strip with periodic boundary conditions is given by the sum over all $2^{L_xL_y}$ spin configurations,
\begin{equation}
\mathcal{Z}
=
\sum_{\{\boldsymbol{\sigma}\}}
\exp\Biggl[
\sum_{i=1}^{L_x}\sum_{j=1}^{L_y}
\left(K_x\,\sigma_{i,j}\sigma_{i+1,j}+K_y\,\sigma_{i,j}\sigma_{i,j+1}
\right)
\Biggr],
\end{equation}
where $K_x$ and $K_y$ are the complex-conjugate couplings along the $x$ and $y$ directions, respectively.
Here, $\sigma_{i,j}=\pm1$, with periodic boundary conditions $\sigma_{L_x+1,j}=\sigma_{1,j}$ and $\sigma_{i,L_y+1}=\sigma_{i,1}$. Taking $\boldsymbol{\sigma}_j=(\sigma_{1,j},\ldots,\sigma_{L_x,j})$
as a slice configuration, the transfer matrix connects two consecutive slices $\boldsymbol{\sigma}_j$ and $\boldsymbol{\sigma}_{j+1}$. The partition function can then be written as $\mathcal{Z}=\operatorname{Tr}(\mathbf{T}^{L_y})$~\cite{Meisinger2013,Baxter1982}.

With the intra-slice weights symmetrically distributed between two consecutive slices, the transfer matrix factorizes as
\begin{equation}
\mathbf{T}=\mathbf{W}^{1/2}\mathbf{V}\mathbf{W}^{1/2},
\end{equation}
where
\begin{equation}
W_{\boldsymbol{\sigma},\boldsymbol{\sigma}'}
=\delta_{\boldsymbol{\sigma},\boldsymbol{\sigma}'}\exp\left(K_x\sum_{j=1}^{L_x}\sigma_j\sigma_{j+1}
\right),
\end{equation}
and
\begin{equation}
V_{\boldsymbol{\sigma},\boldsymbol{\sigma}'}=\exp\left(K_y\sum_{j=1}^{L_x}\sigma_j\sigma_j'\right),
\end{equation}
with $\sigma_{L_x+1}=\sigma_1$~
\cite{10.1093/oso/9780198788102.003.0006,PhysRevE.66.057103}. This factorization ensures that $\mathbf{T}$ is complex
symmetric, $\mathbf{T}^{T}=\mathbf{T}$, but does not restore Hermiticity. Since the couplings are complex, complex symmetry does not imply Hermiticity, and in general $\mathbf{T}^{\dagger}\neq\mathbf{T}$. This marks a fundamental
distinction from the generalized Baxter--Wu model, whose transfer matrix is Hermitian and therefore has a purely real spectrum~\cite{ffsy-k96l}.

The global spin-flip symmetry $\sigma_i\to-\sigma_i$ allows the state space to be decomposed into even and odd spin-flip sectors. The corresponding transfer matrices are
\begin{equation}
T^{\pm}_{a,b}=T_{a,b}\pm T_{a,\bar{b}},
\qquad
a,b=1,\ldots,2^{L_x-1},
\end{equation}
where $\bar{b}$ denotes the spin-flipped partner of state $b$. This decomposition reduces the original $2^{L_x}\times2^{L_x}$ transfer matrix to two independent $2^{L_x-1}\times2^{L_x-1}$ matrices, corresponding to the even ($+$) and odd ($-$) spin-flip sectors~\cite{Bernaschi2024}. The dominant eigenvalue $\lambda_{\max}$ is extracted from the transfer-matrix spectrum and is used to compute the finite-size free energy. Furthermore, the crossings of its imaginary part for different strip widths are used to locate the finite-size spectral transition points, which are then extrapolated to the thermodynamic limit to identify the $\mathcal{PT}$-symmetry-breaking boundary $K_{\rm PT}$.

\begin{figure}
    \centering
    \includegraphics[width=0.5\textwidth]{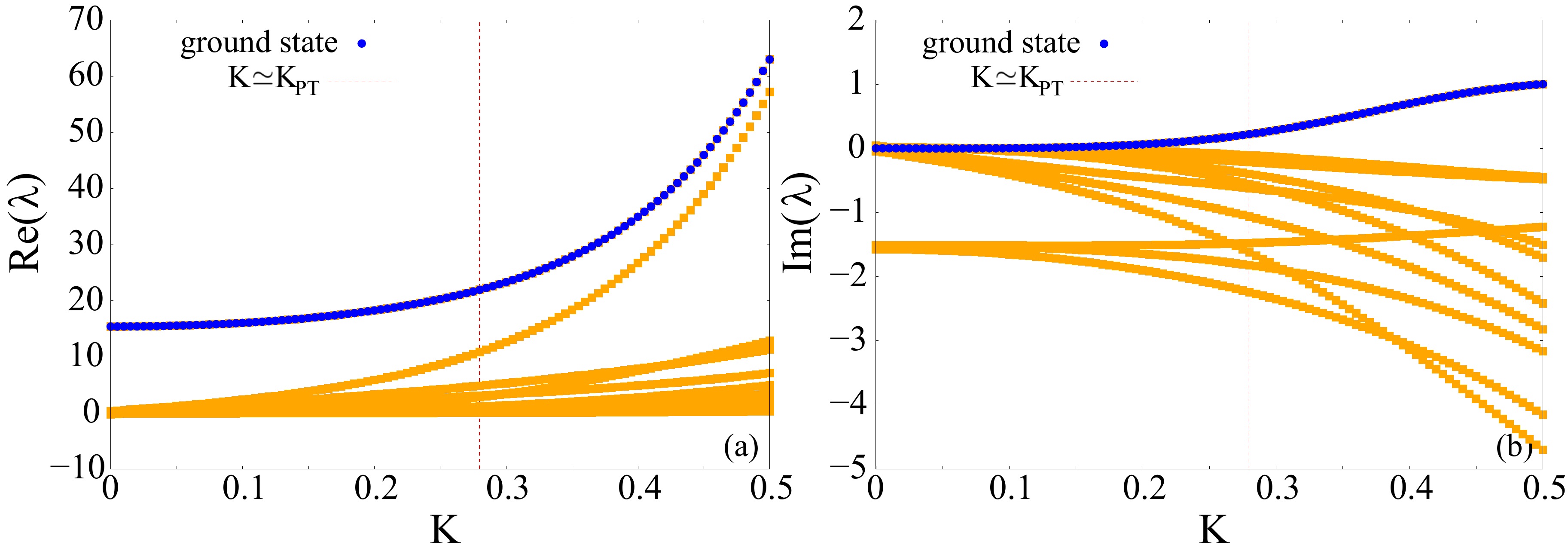}
    \caption{\label{fig:S4}Transfer-matrix eigenvalue spectrum of the CCI model at fixed $\phi=0.1$. (a) Real and (b) imaginary parts of the full eigenvalue spectrum for $L_x=4$. The dominant eigenvalue $\lambda_{\max}$ is highlighted in blue, and the red dashed lines indicate the spectral transition point $K_{\rm PT}$.}
\end{figure}

To provide a complete picture of the spectral structure, we show the full transfer-matrix eigenvalue spectrum at fixed $\phi=0.1$ and $L_x=4$ in Fig.~\ref{fig:S4}. As $K$ increases, the spectral branches exhibit complex evolution. Notably, because the finite strip geometry prevents strict finite-size $\mathcal{PT}$ symmetry, some subdominant eigenvalues may acquire nonzero imaginary parts below $K_{\rm PT}$. However, in the thermodynamic limit, the free energy and macroscopic thermodynamic properties are governed solely by the dominant eigenvalue $\lambda_{\max}$. Therefore, we rigorously define the spectral transition boundary $K_{\rm PT}$ based on the transition of $\lambda_{\max}$ from real to complex, as shown in Fig.~4 of the main text.

\subsection{Mapping to a one-dimensional quantum Ising chain}
To establish a connection between the two-dimensional CCI model and a one-dimensional non-Hermitian quantum spin chain, we consider an approximate quantum Hamiltonian representation based on the transfer-matrix formulation. Taking $K_x=K+i\phi$ as the intra-slice coupling and $K_y=K-i\phi$ as the inter-slice coupling, the latter gives rise to an effective transverse field through the standard dual-coupling transformation. Within this approximation, the transfer matrix can be expressed as
\begin{equation}
\mathbf{T}\simeq
\exp\left[K_x\sum_{i=1}^{M}\hat{\sigma}_i^z\hat{\sigma}_{i+1}^z
+\tilde K_y\sum_{i=1}^{M}\hat{\sigma}_i^x\right],
\end{equation}
where the dual coupling satisfies
\begin{equation}
\tanh\tilde K_y=e^{-2K_y},
\qquad
\tilde K_y=\frac{1}{2}\ln\coth K_y.
\end{equation}
The resulting effective one-dimensional quantum Hamiltonian
$H_Q$, defined through $\mathbf{T}\simeq e^{-\tau H_Q}$
with $\tau=1$, takes the form
\begin{equation}
\begin{aligned}
H_Q
&=-K_x\sum_{i=1}^{M}\hat{\sigma}_i^z\hat{\sigma}_{i+1}^z
-\tilde K_y\sum_{i=1}^{M}\hat{\sigma}_i^x\\
&=-(K+i\phi)\sum_{i=1}^{M}
\hat{\sigma}_i^z\hat{\sigma}_{i+1}^z
-\frac{1}{2}\ln\coth(K-i\phi)
\sum_{i=1}^{M}\hat{\sigma}_i^x.
\end{aligned}
\end{equation}

Using the identity
\begin{equation}
\coth(K-i\phi)
=\frac{\sinh 2K+i\sin 2\phi}
{\cosh 2K-\cos 2\phi},
\end{equation}
we decompose the complex dual coupling as
\begin{equation}
\tilde K_y=h_y+i\gamma_y.
\end{equation}
Its real part is
\begin{equation}
h_y=\frac{1}{4}\ln\left[\frac{\cosh 2K+\cos 2\phi}{\cosh 2K-\cos 2\phi}\right],
\end{equation}
while its imaginary part is
\begin{equation}
\gamma_y=\frac{1}{2}
\arctan\left(\frac{\sin 2\phi}{\sinh 2K}\right),
\end{equation}
where the branch of the complex logarithm is chosen consistently with the parameter range under consideration.

Substituting $K_x=K+i\phi$ and $\tilde K_y=h_y+i\gamma_y$ into the Hamiltonian, we obtain
\begin{equation}
\begin{aligned}
H_Q={}&-K\sum_{i=1}^{M}
\hat{\sigma}_i^z\hat{\sigma}_{i+1}^z
-i\phi\sum_{i=1}^{M}
\hat{\sigma}_i^z\hat{\sigma}_{i+1}^z\\
&-h_y\sum_{i=1}^{M}\hat{\sigma}_i^x
-i\gamma_y\sum_{i=1}^{M}\hat{\sigma}_i^x.
\end{aligned}
\end{equation}
The resulting effective Hamiltonian describes a non-Hermitian transverse-field Ising chain with complex Ising interactions and a complex transverse field. In the limit $\phi=0$, both couplings become real, recovering the Hermitian transverse-field Ising form.
This approximate mapping provides a connection between the CCI model and non-Hermitian quantum spin chains.

\section{Central charge from finite-size free-energy scaling}
\label{C-c}
We extract the central charge from the finite-size correction to the transfer-matrix free energy. The anisotropic Ising transfer matrix admits an exact free-fermion representation. Introducing the dual coupling $K_x^*$ through the Kramers--Wannier relation~\cite{PhysRev.60.252},
\begin{equation}
\sinh(2K_x)\sinh(2K_x^*)=1,
\end{equation}
the single-particle dispersion is given by~\cite{RevModPhys.36.856}
\begin{equation}
\begin{aligned}
\cosh\gamma(q)=\cosh(2K_x^*)\cosh(2K_y)-\sinh(2K_x^*)\sinh(2K_y)\cos q.
\end{aligned}
\end{equation}
On the analytically continued critical line, $K_x^*=K_y$. Choosing the branch continuously connected to the real ferromagnetic Ising model gives
\begin{equation}
\gamma(q)=2\operatorname{arsinh}\!
\left[\sinh(2K_y)\sin\frac q2\right].
\end{equation}
For the Neveu--Schwarz (NS) momenta
\begin{equation}
q_r=\frac{(2r+1)\pi}{L},
\qquad r=0,\ldots,L-1,
\end{equation}
the corresponding vacuum eigenvalue satisfies
\begin{equation}
\ln\lambda_{\rm NS}(L)=
\frac L2\ln[2\sinh(2K_x)]
+\frac12\sum_{r=0}^{L-1}\gamma(q_r).
\end{equation}
All logarithms and inverse hyperbolic functions are evaluated on branches continuously connected to the real ferromagnetic Ising model. For the critical parameters considered here, the NS vacuum branch coincides with the spectral-radius branch of the finite-width transfer matrix. We denote the corresponding dominant eigenvalue by $\lambda_0(L)$.

Taking the real part of the logarithm and replacing the momentum sum by an integral yields the thermodynamic-limit real free-energy density $f_{R,\infty}$ in Eq.~(6) of the main text. Combining this limit with the finite-size eigenvalue gives Eq.~(7) of the main text, the exact anisotropy-corrected free-fermion prediction shown by the solid lines in Fig.~3(f). This finite-$L$ expression contains no fitting parameters.

To derive the leading finite-size correction, we expand the dispersion near $q=0$:
\begin{equation}
\gamma(q)=v_xq+O(q^3),
\qquad v_x=\sinh(2K_y).
\end{equation}
Applying the midpoint Euler--Maclaurin formula to the NS momentum sum gives
\begin{equation}
\begin{aligned}
\operatorname{Re}\sum_{r=0}^{L-1}\gamma(q_r)
={}&\frac{L}{2\pi}\operatorname{Re}
\int_0^{2\pi}\gamma(q)\,dq\\
&+\frac{\pi\operatorname{Re}v_x}{6L}
+O(L^{-3}),
\end{aligned}
\end{equation}
where $\gamma'(0)=v_x$ and $\gamma'(2\pi)=-v_x$. Substitution into the free-energy expression yields
\begin{equation}
f_R(L)=f_{R,\infty}
-\frac{\pi\operatorname{Re}v_x}{12L^2}
+O(L^{-4}).
\end{equation}

For the CCI critical couplings $K_x=K_c+i\phi$ and $K_y=K_c-i\phi$, the analytically continued critical condition implies
\begin{equation}
\sinh^2(2K_c)+\sin^2(2\phi)=1.
\end{equation}
On the ferromagnetic branch considered here, $\sinh(2K_c)=\cos(2\phi)$. Consequently,
\begin{equation}
\begin{aligned}
\operatorname{Re}v_x
&=\operatorname{Re}\sinh(2K_c-2i\phi)\\
&=\sinh(2K_c)\cos(2\phi)\\
&=\cos^2(2\phi).
\end{aligned}
\end{equation}

For a two-dimensional critical system with periodic transverse boundary conditions, conformal field theory relates the leading finite-size free-energy correction to the central charge~\cite{PhysRevLett.56.742,PhysRevLett.56.746}. Comparing the standard isotropic result with the expansion above, the coefficient obtained without correcting for anisotropy is
\begin{equation}
c_{\rm fit}=c\operatorname{Re}v_x
=c\cos^2(2\phi).
\end{equation}
Using the anisotropy-corrected definition of $c_{\rm eff}(L)$ in Eq.~(8) of the main text, the finite-size expansion gives
\begin{equation}
c_{\rm eff}(L)=c+O(L^{-2}).
\end{equation}
Thus, extrapolating $c_{\rm eff}(L)$ in $1/L_x^2$ recovers the universal Ising central charge $c=1/2$.

\end{document}